\documentclass{iopart}
\usepackage{iopams}  
\jl{32}
\def\={\equiv} 
\def\div{\nabla\cdot } 
\def\pl{\partial}

\newcommand{\bib}{\bibitem}

\newcommand{\lab}{\label}

\newcommand{\la}{\langle\,}

\newcommand{\ra}{\,\rangle}

\newcommand{\3}[1]{{\boldsymbol #1}}
\newcommand{\bb}[1]{{\boldsymbol{\bar #1}}}

\newcommand{\6}[1]{_{\scriptscriptstyle#1}}

\newcommand{\8}{\infty}

\def\c{\chi}

\def\m{\mu} 
\def\n{\nu}
\def\o{\omega} 
\def\p{\pi}

\def\y{\psi}

\def\P{\Pi}
 
\def\S{\Sigma} 

\def\Y{\Psi}

\newcommand{\db}{{\,{\rm d}\kern-.9ex {^-}}\!}
\newcommand{\dir}{{\pl\kern-1.2ex {/}}}
\newcommand{\dd}{{\,\rm d}}

\newcommand{\curl}{\nabla\times}

\newcommand{\ie}{{\it ie, }}

\newcommand{\imp}{\ \Rightarrow\ }

\newcommand{\ir}{\int_{-\infty}^\infty} 

\newcommand{\lra}{\leftrightarrow}

\newcommand{\plra}{\pl^{\kern-1.25ex^\lra}}
\newcommand{\qq}{\quad} 
\newcommand{\qqq}{\qquad} 
\newcommand{\re}{{\,\rm Re}\  }   
\newcommand{\rr}[1]{{{\mathbb R}^{#1}}}

\newcommand{\sgn}{{\,\rm Sgn \,}}
\newcommand{\sh}[1]{\hskip#1ex}

\newcommand{\orr}{{(\3r)}}
\newcommand{\ort}{{(\3r,t)}}

\def\Xint#1{\mathchoice
   {\XXint\displaystyle\textstyle{#1}}%
   {\XXint\textstyle\scriptstyle{#1}}%
   {\XXint\scriptstyle\scriptscriptstyle{#1}}%
   {\XXint\scriptscriptstyle\scriptscriptstyle{#1}}%
   \!\int}
\def\XXint#1#2#3{{\setbox0=\hbox{$#1{#2#3}{\int}$}
     \vcenter{\hbox{$#2#3$}}\kern-.5\wd0}}

\def\ppint{\Xint-}

\def\HB{\hfill\break}

\begin{document}

\title{Helicity, polarization, and Riemann-Silberstein vortices}

\author{Gerald Kaiser}

\address{Virginia Center for Signals and Waves \\ www.wavelets.com}

\begin{abstract}
\it Riemann-Silberstein \rm (RS) \it vortices \rm have been defined as surfaces in spacetime where the complex form of a free electromagnetic field given by $\3F=\3E+i\3B$  is null ($\3F\cdot\3F=0$), and they can indeed be interpreted as the collective history swept out by moving vortex lines of the field. Formally, the nullity condition is similar to the definition of  \it  C-lines \rm associated with a monochromatic electric or magnetic field, which are curves in space where the polarization ellipses degenerate to circles. However, it was noted that RS vortices of monochromatic fields generally oscillate at optical frequencies and are therefore unobservable, while electric and magnetic C-lines are steady. Here I show that under the additional assumption of having definite \it helicity, \rm RS vortices are not only steady but they coincide with \it both \rm sets of C-lines, electric and magnetic.  The two concepts therefore become one for waves of definite frequency and helicity.  Since the definition of RS vortices is relativistically invariant while that of C-lines is not, it may be useful to regard the vortices as a wideband generalization of C-lines for waves of definite helicity.
\end{abstract}

\pacs{03.50.De, 41.85.-p,  42.25.Ja}

\submitted

\section{Polarization singularities of monochromatic vector waves}

Monochromatic electric and magnetic fields generically contain lines (curves)  where the polarization becomes either purely linear or purely circular. This important fact was predicted by Nye and Hajnal \cite{NH87} and observed by Hajnal \cite{H90}; for excellent and up-to-date accounts, see \cite{BD01,D01,N99}.  Though a full analysis employs sophisticated methods from singularity theory and topology,  the basics are easily understood.  A real, time-harmonic vector field in $\rr3$ has the form
\begin{equation}\lab{E}
\3E(\3r,t)=\re \{e^{-i\o t}\3E_\o(\3r) \}, \qqq \3E_\o(\3r)=\3P(\3r)+i\3Q(\3r),
\end{equation}
which sweeps out an ellipse $\5E_\3r$  at $\3r$ as time flows.  This defines a \it  field \rm of polarization ellipses in $\rr3$. In general, the major and minor axes of $\5E_\3r$ together with their cross product define a unique orthogonal frame at $\3r$ and, taking into account the arrow of time, this gives a variable field of \it  oriented \rm frames in $\rr3$ containing much information about $\3E$. However, the construction breaks down at points where the ellipse becomes a circle or collapses to a line, since no \it  unique \rm frames exist there.  Points of circular polarization are characterized by the two real equations  $\3P\cdot\3P=\3Q\cdot\3Q$ and $\3P\cdot\3Q=0$, or more compactly by the single complex equation
\begin{equation}\lab{ps}
\Y\6E(\3r)\=\3E_\o(\3r)^2=0, \ \hbox{\ where\  } \ \3E_\o^2\=\3E_\o\cdot\3E_\o.
\end{equation}
For general $\3r$, the complex function $\Y\6E(\3r)$ is called the \it  polarization scalar \rm associated with $\3E$.  Generic solutions of $\Y\6E(\3r)=0$ have dimension $3-2=1$, and such curves, called \it  C-lines, \rm  are clearly stationary.  Points of linear polarization are characterized by $\bb E_\o\times\3E_\o=2i\3P\times\3Q=\30$ and also occur generically along curves, called \it  L-lines. \rm

Applied to a monochromatic electromagnetic field $(\3E, \3B)$, this gives electric and magnetic C-lines and L-lines which are in general independent in spite of the fact that the two fields are coupled by Maxwell's equations.

\section{Riemann-Silberstein vortices}

A promising relativistic construction \it  analogous \rm to the above was recently initiated by Iwo and Zofia Bialynicki-Birula \cite{BB03}.  The complex structure of $\3E_\o$ in  \eref{E} is related to the frequency by   
\begin{equation}\lab{Eo}
 \3E_\o(\3r)=\3E(\3r, t_n)+i\3E(\3r, t_n+\p/2\o),
 \qq t_n=2\p n/\o.
\end{equation}
Not only is this nonrelativistic, but every choice of frequency gives a different complex structure.  On the other hand, the complex structure in  \cite{BB03} is based on a symmetry common to \it all \rm solutions of the source-free Maxwell equations (with $c=1$), namely the mapping acting on solutions by
\begin{equation}\lab{duality}
\5J(\3E(\3r,t),\3B(\3r,t))=(-\3B(\3r,t),\3E(\3r,t)).
\end{equation}
Since $\5J^2$ is minus the identity, this suggests defining the complex field
\begin{equation}\lab{F}
\3F(\3r,t)=\3E(\3r,t)+i\3B(\3r,t),
\end{equation}
on which \eref{duality} is represented as multiplication by $i$. (All that follows is invariant under $i\to -i$ as there is no intrinsic difference between these two eigenvalues of $\5J$.) For historical reasons \cite{S14}, Bialynicki-Birula \cite{B96} calls $\3F$ the \it  Riemann-Silberstein \rm (RS) vector. It also comes up naturally in the theories of spinors and twistors \cite{HT94}.  $\5J$ generates a one-parameter group of \it  duality rotations, \rm  acting on $\3F$ by
\begin{equation}
\3F(\3r,t)\to e^{i\c}\3F(\3r,t), \lab{drot}
\end{equation}
where $\c=\p/2$ gives \eref{duality}. From a foundational viewpoint, electromagnetic duality has traditionally been regarded as an accidental symmetry since it cannot be extended to sources in the absence of magnetic charges. 
However, recent progress in the theory of fundamental interactions suggests that duality is a broken symmetry with deep roots; see \cite{O99}. Since polarization sources exist equally in electric and magnetic form, they can be accommodated in the equations without breaking duality. In \cite{K3,K4}, complex  combinations of polarization sources are used to generate `electromagnetic wavelets.'

From here on we confine attention to sourceless fields, where Maxwell's equations take the simple complex form
\begin{equation}\lab{M}
\div\3F=0, \qq i\pl_t\3F=\curl \3F
\end{equation}
with the coupling between $\3E$ and $\3B$ now represented by the factor $i$, a duality rotation. These equations are relativistic in spite of their appearance,  with Lorentz trans\-form\-ations represented by complex $3\times3$ matrices. (In covariant notation, $\3F$ corresponds to the \it self-dual \rm part $F_{\m\n}-iF^*_{\m\n}$ of the field tensor.) Even simpler is the interpretation of $\3F$ as a complex $2\times2$ matrix in the  Clifford-Pauli algebra, where Lorentz transformations take on a spinorial form $\3F\to L\3F\7L$; see Baylis  \cite[p 71]{B99}.

It is reasonable to wonder if the RS field has a physical interpretation beyond that of its real and imaginary parts.  Since \eref{M} is essentially an equation of the Weyl type for a massless particle of spin 1, it has been  proposed that $\3F$ can represent a photon wave function as well as a classical Maxwell field; see  \cite{B96}.

Now that we have a complex vector field, we may define the `polarization scalar'
\begin{equation}\lab{PS}
\y(\3r,t)=\3F(\3r,t)^2
\end{equation}
and inquire about the meaning of its zeros, which are generically surfaces in spacetime. Note that $\y(\3r,t)$, unlike the electric and magnetic polarization scalars $\Y\6E(\3r), \Y\6B(\3r)$,   is Lorentz invariant, making the definition of these surfaces invariant. They can be viewed in any reference frame as the collective history of the generic space curves given at time $t$ by $\y(\3r,t)=0$, which are interpreted as \it  vortex lines \rm of the field in \cite{BB03}; hence the surfaces are called \it  RS vortices. \rm 

\section{Helicity}

We want to know whether the similarity between C-lines and RS vortices is more than an analogy. Since C-lines are defined for 
monochromatic fields, consider the frequency decomposition of a general RS field,
\begin{equation}\lab{freq}
\eqalign{
\fl\3F\ort=\ir d\o\ e^{-i\o t} \3F_\o\orr=\3F\6+\ort+\3F\6-\ort\\
\fl\3F\6+\ort=\int_0^\8 d\o\  e^{-i\o t} \3F_\o\orr, \qq
\3F\6-\ort=\int_0^\8 d\o\  e^{i\o t} \3F_{-\o}\orr.
}\end{equation}
Since $\3F$ is complex, its positive and negative frequency parts
$\3F\6\pm$  are independent as Fourier coefficients, \ie they need not  satisfy the reality condition $\3F_{-\o}=\bb F_\o$. The decomposition is 
also invariant under proper Lorentz transformations.  Furthermore, Maxwell's equations \eref{M} become
\begin{equation}\lab{M2}
 \div\3F\6\pm=0,\qq \curl\3F\6\pm=i\pl_t\3F\6\pm,\qq
\curl\3F_\o=\o\3F_\o
\end{equation}
and the last equation implies $\div\3F_\o=0$ for $\o\ne 0$. (Since $\div \3F_0=0$ and $\curl\3F_0=\30$, $\3F_0\orr$ is constant; we assume $\3F_0\orr\=\30$, dealing only with propagating waves.)

Equations \eref{M2} show that $\3F\6\pm$ are independent solutions of Maxwell's equations. Thus, in particular, $\3F_{\pm\o}$ are independent not only as Fourier coefficients but also as solutions.  On the other hand, 
the real and imaginary parts of $\3F\6\pm$ are still coupled by the factor $i$ as in \eref{M}, while those of  $\3F_\o$ and $\3F_{-\o}$ are  uncoupled.  This just one of the benefits of the RS representation.  The real fields  are given by
\begin{equation}\lab{EB}
\eqalign{
\3E\ort=\re\3E\6+\ort,\qq \3E\6+\ort=\3F\6+\ort+\bb F\6-\ort\\
\3B\ort=\re\3B\6+\ort,\qq i\3B\6+\ort=\3F\6+\ort-\bb F\6-\ort.
}\end{equation}
Note that $\3E\6+, \3B\6+$ are the positive frequency parts of $\3E, \3B$, and each has contributions from both components $\3F\6\pm$ of $\3F$.  Equation \eref{EB} is a wideband version of \eref{E}, reducing to it in the monochromatic case:
\begin{equation}\lab{EB2}
\eqalign{
\3F\ort=e^{-i\o t}\3F_{\o}\orr+e^{i\o t}\3F_{-\o}\orr\\
\3E\6+\ort=e^{-i\o t}\3E_\o\orr\  \  \hbox{where}\  \  \ 
\3E_\o\orr=\3F_\o\orr+\bb F_{-\o}\orr\\
 \3B\6+\ort=e^{-i\o t}\3B_\o\orr\  \  \hbox{where}\  \ 
i\3B_\o\orr=\3F_\o\orr-\bb F_{-\o}\orr.
}\end{equation}
Now consider a single plane-wave solution,
\begin{equation}
\fl\sh8 \3F\ort=e^{i\3k\cdot\3r-i\o t}\3f\imp i\3k\times\3F=\o\3F
\imp \3F^2=0, \  \  \o=\pm|\3k|.
\end{equation}
Plane waves do not have generic vortices since $\3F^2$ vanishes everywhere. If $\o>0$, then $\{\3E, \3B, \3k\}$ form a right-handed orthogonal frame. The fields spin clockwise around $\3k$ (viewed from the rear) as they propagate, and we say that $\3F$ has \it positive helicity. \rm  If $\o<0$, they form a left-handed frame, spin counterclockwise, and $\3F$ has negative helicity.  The helicity of a plane wave is just the sign of the frequency, but we can extend it to an operator $\5S$ acting on \it all \rm solutions by
\begin{equation}
\5S\3F_\o=(\sgn\o)\,\3F_\o \qqq \5S(\3F\6++\3F\6-)=\3F\6+-\3F\6-.
\end{equation}
For general solutions, this is just $i$ times the Hilbert transform,
\begin{equation}
\5S\3F(\3r, t)=\frac i\p\ppint \frac{\dd t'}{t'-t}\, \3F(\3r, t'),
\end{equation}
where the integral denotes the principal value. Thus, every propagating solution  decomposes uniquely and (Lorentz-) invariantly into eigenstates $\3F\6\pm$ of positive and negative helicity. (Recall that $\3F_0=\30$; otherwise $\5S$ would be undefined on $\3F_0$.)

To compare vortices and C-lines, return to the monochromatic solution \eref{EB2}. Its `polarization scalar' is
\begin{equation}\lab{polscalar}
\y=e^{-2i\o t}\3F_\o^2+2\3F_\o\cdot\3F_{-\o}+e^{2i\o t}\3F_{-\o}^2
\end{equation}
showing that generic RS vortices oscillate at the optical frequency $2\o$ and are therefore unobservable. This is noted by Berry  \cite{B03}, who therefore considers the \it  time average \rm 
\begin{equation}\lab{time}
\fl \Y(\3r)\=\la \y(\3r,t)\ra_{\rm time}=2\3F_\o(\3r)\cdot\3F_{-\o}(\3r)
=\frac12(\3E\6++i\3B\6+)\cdot(\bb E\6++i\bb B\6+).
\end{equation}
The generic curves where $\Y(\3r)=0$ are thus interpreted as `fuzzy' vortex lines due to the fluctuations in $\y$. See also Dennis 
 \cite[pp 115--16]{D01}.
 
Here I propose a different approach.  Since the equations for the real and imaginary parts of $ \3F_\o$ are uncoupled, they may be considered individually. \Eref{M2} states that their   vorticites  are everywhere proportional to themselves by the constant factor $\o$, hence their flow lines are \it twisted \rm everywhere to the same degree $|\o|$, in a positive or negative sense depending on the helicity.  Such fields were first studied by Beltrami in fluid mechanics and now play important roles in plasma physics and many other areas. Equation  \eref{M2} shows that  the positive and negative helicity components in $\3F$ engage in a tug of war, which explains why RS fields of mixed helicity have unstable vortices as seen in the monochromatic example above.

Given a general wave $\3F$, the above argument suggests that we consider separately the vortices of its helicity components $\3F\6\pm$.  Note that $\3F\6\pm$ are independent solutions, hence they could be separated  in principle. (In fact, helicity is also a quantum observable, so the field could be prepared as a helicity eigenstate to begin with.)  It would be interesting to 
have an experimental confirmation of the above considerations.  

For the monochromatic case with $\o>0$, \eref{EB2} gives
\begin{equation}\lab{RSC}\eqalign{
\fl\3F\ort=e^{-i\o t}\3F_\o\orr &\imp\3E\6+\ort=i\3B\6+\ort=e^{-i\o t}\3F_\o(\3r)\\
\fl &\imp\3E(\3r, t)=\3B(\3r, t-p/4), \qq  p=2\p/\o.
}\end{equation}
Thus $\3E$ trails $\3B$ by a quarter-period as expected, and the two fields generate  identical polarization ellipse fields. The polarization scalars are related by
\begin{equation}
\y(\3r,t)=e^{-2i\o t}\Y\6E(\3r)=-e^{-2i\o t}\Y\6B(\3r).
\end{equation}
Therefore  \it  the C-lines of $\3E$, the  C-lines of $\3B$, and the vortices of $\3F$  all  coincide. \rm  The same goes for negative $\o<0$, where $\3B$ trails $\3E$  by a quarter-period.

It is amusing to consider the above system from a moving reference frame. Since $\y$ is relativistically invariant, the vortices remain vortices. But they are no longer C-lines in the usual sense as the field is no longer monochromatic. This suggests viewing the vortices of \it arbitrary \rm  helicity eigenstates as generalized C-lines.  Of course, this view cannot accommodate all C-lines since any invariant definition must take into account both $\3E$ and $\3B$ whereas C-lines are generally used to study electric and magnetic polarization singularities individually.

\ack

I thank Michael Berry, Iwo Bialynicki-Birula  and Mark Dennis for a number of informative discussions during the Singular Optics 2003 Workshop, where I first encountered this fascinating subject, and later by email. I also thank Dr.~Arje Nachman for supporting my work through AFOSR Grant  \#F49620-01-1-0271.

\Bibliography{20}

\bib{B99} Baylis W E 1999,  \it  Electrodynamics: A Modern
Geometric Approach. \rm (Boston: Birkh\"auser,  Progress in Mathematical Physics vol~17)
 
\bib{BD01}  Berry M V and Dennis M 2001,
Polarization singularities in isotropic random vector waves. 
\it Proc R Soc Lond \rm A \bf 457 \rm 141--55.
\ www.phy.bris.ac.uk/research/theory/Berry/

\bib{B03}  Berry M V 2003, Riemann-Silberstein vortices for paraxial waves. Submitted to \it J Optics A: Pure Appl Opt \rm (special issue on Singular Optics) July 2003

\bib{B96}   Bialynicki-Birula I 1996, Photon wave function. \it  Progress in Optics \rm vol~36 ed E Wolf  (Amsterdam: North-Holland).
\    www.cft.edu.pl/~birula/publ.html

\bib{BB03}   Bialynicki-Birula I and Bialynicka-Birula Z  2003,  Vortex lines of the electromagnetic field.  \PR A \bf 67 \rm 062114.
\    www.cft.edu.pl/~birula/publ.html 

\bib{D01}  Dennis M 2001, \it  Topological Singularities in Wave Fields, \rm  Ph D thesis  Physics  University of Bristol.
\  www.phy.bris.ac.uk/staff/dennis\_\,mr.htm \rm

\bib{H90}   Hajnal J V 1990,  Observation of singularities in the electric and magnetic fields of freely propagating microwaves. 
\it  Proc R Soc Lond \rm A \bf 430 \rm 413--21

\bib{HT94}   Huggett S A and  Todd, K P 1994,  \it   An Introduction to Twistor Theory  \rm  (Cambridge: Cambridge University Press)

\bib{K3} Kaiser G  2003,   Physical wavelets and their sources:
Real physics in complex space-time. \HB 
Topical Review,  \JPA \bf 36 \rm no~30   R291--R338. \HB  www.iop.org/EJ/toc/0305-4470/36/30 

\bib{K4} G Kaiser, \it Making electromagnetic wavelets. \rm
http://arxiv.org/abs/math-ph/math-ph/0402006

\bib{N99}  Nye  J F 1999,  \it  Natural Focusing and Fine Structure of Light. \rm (Bristol: Institute of Physics Publishing)

\bib{NH87}  Nye J F  and Hajnal J V 1987, The wave structure of monochromatic electromagnetic radiation.  \it  Proc R Soc Lond \rm  A \bf 409 \rm 21--36

\bib{O99}  Olive D 1999, Introduction to duality, in \it  Duality and Supersymmetric Theories, \rm eds D Olive and P C West
(Cambridge: Cambridge University Press).  See also  
\  arxiv:hep-th/9508089

\bib{S14} Silberstein L 1914, \it The Theory of Relativity \rm (London: MacMillan and Company)

\endbib

\end{document}